\documentclass[aps,twocolumn,floats,prd,nofootinbib,superscriptaddress,showkeys,10pt]{revtex4-2}

\usepackage{graphicx} 
\usepackage{amsfonts,amsmath,amssymb} 
\usepackage{hyperref}

\begin{document}
\title{Universality in static spherically symmetric solutions of $f(R)$ gravity}

\author{V.~I.~\surname{Zhdanov}}
\email{valery.zhdanov@knu.ua}
\affiliation{Taras Shevchenko National University of Kyiv,  Kyiv 01601, Ukraine}

\date{May 2025}

\begin{abstract}
 $f(R)$ gravity is a well-known modification of General Relativity, which is used in various models of dark matter and dynamical dark energy. We study static spherically symmetric (SSS) asymptotically flat configurations of the $f(R)$ gravity in the Einstein frame for three  known  scalaron potentials with   different asymptotic properties.  The main attention is paid to the case of astrophysically relevant configuration mass $M$ and scalaron mass $\mu$ greater than several $meV$, according to existing experimental constraints. This means very large values of the dimensionless (in geometrized units) parameter $M\mu$, leading to specific  properties of the SSS  solutions. We consider a sufficiently large size of a  scalarization region $r_0\gg r_g$, where the space-time metric is essentially different from the Schwarzschild one. It turns out that the scalaron field has universal behavior regardless of $M\mu$ and $r_0$ and is practically the same for different scalaron potentials.
Asymptotic parameters of the metric near the naked singularity at the center of the SSS configuration are obtained analytically for all the models. The test body circular orbits are discussed in view of observational signatures,  which can distinguish these  configurations from a regular black hole. 
\end{abstract}


\maketitle%


\section{Introduction} \label{Introduction}
\textbf{The $f(R)$ gravity is a family of theories generalizing Einstein's  General Relativity} (see, e.g., \cite{Sotiriou:2008rp, DeFelice:2010aj, Nojiri:2017ncd} for a review). It is used in inflationary models of very early Universe, in dark matter models etc \cite{Starobinsky:1980te, Vilenkin1985, Shtanov-Sahni2023, Odintsov-2023, Capozziello:2006uv, Nojiri:2008nt, Cembranos:2008gj, Cembranos:2015svp, Corda:2011aa, Katsuragawa:2016yir, Katsuragawa:2017wge, Yadav:2018llv, Parbin:2020bpp, Shtanov:2021uif, Shtanov:2022xew}. The gravitational action in these theories is 
\begin{equation} \label{action}
S_g=-\frac{1}{2\kappa} \int d^{4} x \, \sqrt{-g} \,f(R)   \, , 
\end{equation}
where   $R$ is the scalar curvature\footnote{the metric signature is $(+,-,-,-)$, $R^\alpha_{\,\,\beta\gamma\delta}=\partial_\gamma\Gamma^\alpha_{\beta\delta}-...\,\,R_{\mu\nu}=R^\alpha_{\,\,\mu\alpha\nu}$..;  $c=1$.}. Here, the cosmological constant and non-gravitational matter fields are not included. Typically, the first two terms of expansion in powers of $R$ can be written as  $f(R)=R-R^2/(6\mu^2)$, where $\mu$ is the scalaron mass in geometrized units. 
The action (\ref{action}) leads to the fourth order  dynamical equations   for the physical metric $g_{\mu\nu}$ (Jordan frame). 
However, for a number of  $f(R)$ versions the problem can be reduced to the usual Einstein equations  with respect to some  new metric $\hat{g}_{\mu\nu}$ (Einstein frame) by means of   conformal transformation \cite{Sotiriou:2008rp, DeFelice:2010aj, Nojiri:2017ncd} 
\begin{equation}\label{conformal_xi} \hat{g}_{\mu\nu} = e^{2\xi} g_{\mu\nu}\, ,
\end{equation}
where the scalar field $\xi$ with mass $\mu$ 
dubbed scalaron\footnote{In fact, the canonical scalaron \cite{Sotiriou:2008rp, DeFelice:2010aj, Nojiri:2017ncd}  is obtained by some rescaling of $\xi$. However, following \cite{ZhdStSht2024}, we use the dimensionless $\xi$.} incorporates additional degree of freedom. The scalaron   obeys a  nonlinear wave equation with a self-interaction potential $W(\xi)$ that can be introduced parametrically as follows \cite{ZhdStSht2024}:
\begin{equation} \label{SF-potential}
 e^{2\xi}=f'(u)\,,\quad 
 W(\xi)=\frac12  e^{-4\xi}\left[f(u)-f'(u)u\right]\,.
 \end{equation}

One of the simplest scalaron  potentials arises in the Starobinsky quadratic $f(R)$ model (SR2) \cite{Starobinsky:1980te, Vilenkin1985}. Here, scalaron potential $W(\xi)$ has an infinite plateau for $\xi \to \infty$. This model is one of the most favorable ones in view of the Plank data \cite{Planck:2018jri} (see, e.g., \cite{Shtanov-Sahni2023} and references therein). 
 On the other hand, it turns out \cite{Shtanov-Sahni2023} that many  f(R) theories lead to physically interesting `tabletop' (TT) or `hilltop' (HT) potentials, which  decrease to zero when $\xi\to\infty$.

A number of works on the $f(R)$ gravity are devoted to static spherically symmetric (SSS) configurations  \cite{Cruz-Dombriz2009,Bhattacharya2016,Canate_2018,Nashed2020,Nashed2021, Hernandez2020,ZhdStSht2024}. Most of them investigate possibility of black holes. It is obvious that for $\xi\equiv 0$, when $f(R)$ theory is equivalent to the General Relativity,  the SSS configuration is described by the Schwarzschild metric.The occurrence of an arbitrarily small but  non-trivial $\xi$ can lead to a qualitatively different solution with naked singularity at the center. This was shown in  \cite{Hernandez2020,ZhdStSht2024} for the SR2 model. 
The appearance of a naked singularity is somewhat alarming in light of the well-known hypothesis of cosmic censorship \cite{Penrose1965,Penrose2002}. This  requires a detailed study of corresponding solutions near the singularity, asymptotic properties etc (see, e.g.,  discussions of this subject in \cite{SSZ2023quasinormal,ZhdStSht2024} and references therein). The question arises whether the qualitative properties of SSS configurations within the TT or HT scalaron potentials will change compared to the SR2 model.

We note that, in case of typical  astrophysical objects, 
 the value $M\mu$ is very large\footnote{In geometrized units $M\mu$ is dimensionless.}. Indeed, it was pointed out \cite{Shtanov:2022xew,Cembranos:2008gj} that the  scalaron mass  $\mu$ must be larger than several millielectron volts in view of existing experimental bounds. A number of other models deal with a considerably larger scalaron masses (e.g., \cite{Starobinsky:1980te}). But even the  relatively  small value of $\mu=4\,\text{meV}$  \cite{Shtanov:2021uif} corresponds to the length scale  $l_{\mu} =\mu^{-1}= 0.3~\text{mm}$, which gives   ${\mu}M_{\odot}\simeq 5\cdot 10^6$. And we have an additional  7$\div$9 orders of magnitude for typical masses of  central objects in active galactic nuclei.

\textbf{In this paper, we study SSS solutions of $f(R)$ gravity models up to $M\mu \sim 10^{20}$, including astrophysically interesting configurational masses (in contrast to the numerical results presented in \cite{Hernandez2020,ZhdStSht2024}, where moderate  $M\mu$ were considered). Under these assumptions, based on semi-analytical estimates and numerical simulations, we identify interesting features of the solutions that turn out to be typical for a broader class of SSS models than SR2 considered in \cite{Hernandez2020,ZhdStSht2024}. To this end, we consider the important $f(R)$ cases found in \cite{Shtanov-Sahni2023} and extend our consideration to nonmonotonic potentials of HT and TT scalarons.}

Section \ref{section:potentials} reviews the  basic relations of $f(R)$ gravity in the Einstein frame for the SSS configurations and presents examples of potentials we deal with, which include the non-monotonic HT and TT models. In Section \ref{section:reduction}, the basic equations are presented in equivalent forms suitable for further investigation and for a numerical integration. In Section \ref{section:transition}, analytical and numerical estimates related to the transition between regions of weak and strong scalaron field in asymptotically flat configurations are obtained. Numerical solutions related to the strong field region are presented in Section \ref{section:numerical} and used in Section \ref{Jordan_frame} to discuss the circular geodesics in the Jordan frame.  In Section \ref{section:discussion}, we sum up the results. Appendix \ref{Appendix_A} concerns the approximate solutions in the weak-field region.

 \section{Scalaron potentials}\label{section:potentials}
The absence of non-gravitational fields in the Jordan frame means that in the Einstein frame we deal with the scalar vacuum, namely, we deal with a non-zero scalar field with some scalaron potential $W(\xi)$  \cite{Sotiriou:2008rp, DeFelice:2010aj, Nojiri:2017ncd}. 
 
 For the SSS space-time in the Einstein frame we use the Schwarzschild representation of the  spacetime interval  squared 
\begin{equation} \label{metric} d\hat s^2=
 \hat g_{\mu\nu}dx^\mu dx^\nu =e^{\alpha } dt^{2} -e^{\beta } dr^{2} -r^{2} dO^2 ,
\end{equation} 
where $r>0$ and $dO^2=d\theta ^{2} +\sin^{2} \theta d\varphi^{2}$ stands for the metric element on the unit sphere.
Equation for scalaron $\xi$ is a usual SSS scalar field equation with potential $W/6$ \cite{ZhdStSht2024}: 
\begin{equation} 	
\label{equation-xi-vacuum}
\frac{d}{dr}\left[r^2 e^{\frac{\alpha-\beta}{2}}\frac{d\xi}{dr}\right]= \frac{r^2}{6} e^{\frac{\alpha+\beta}{2}} W'(\xi)   \,.
\end{equation}
The nontrivial Einstein equations in this case are  \cite{ZhdStSht2024}
\begin{align}
&\frac{\partial}{\partial r} \left[ r \left(e^{-\beta}-1\right) \right]= -
r^2 \left[ 3 e^{-\beta}\left(\frac{\partial\xi}{\partial r}\right)^2+  W(\xi)\right] \,,
    \label{Ein_1-0a} \\[3pt] 	
&{re^{-\beta}\frac{\partial\alpha}{\partial r}+e^{-\beta}-1} = r^2 \left[3 e^{-\beta}\left(\frac{\partial\xi}{\partial r}\right)^2-  W(\xi)\right]\,,
\label{Ein_2-0a}
\end{align}

\begin{figure}[h!]  	\centering
\includegraphics[width=0.55 \textwidth]{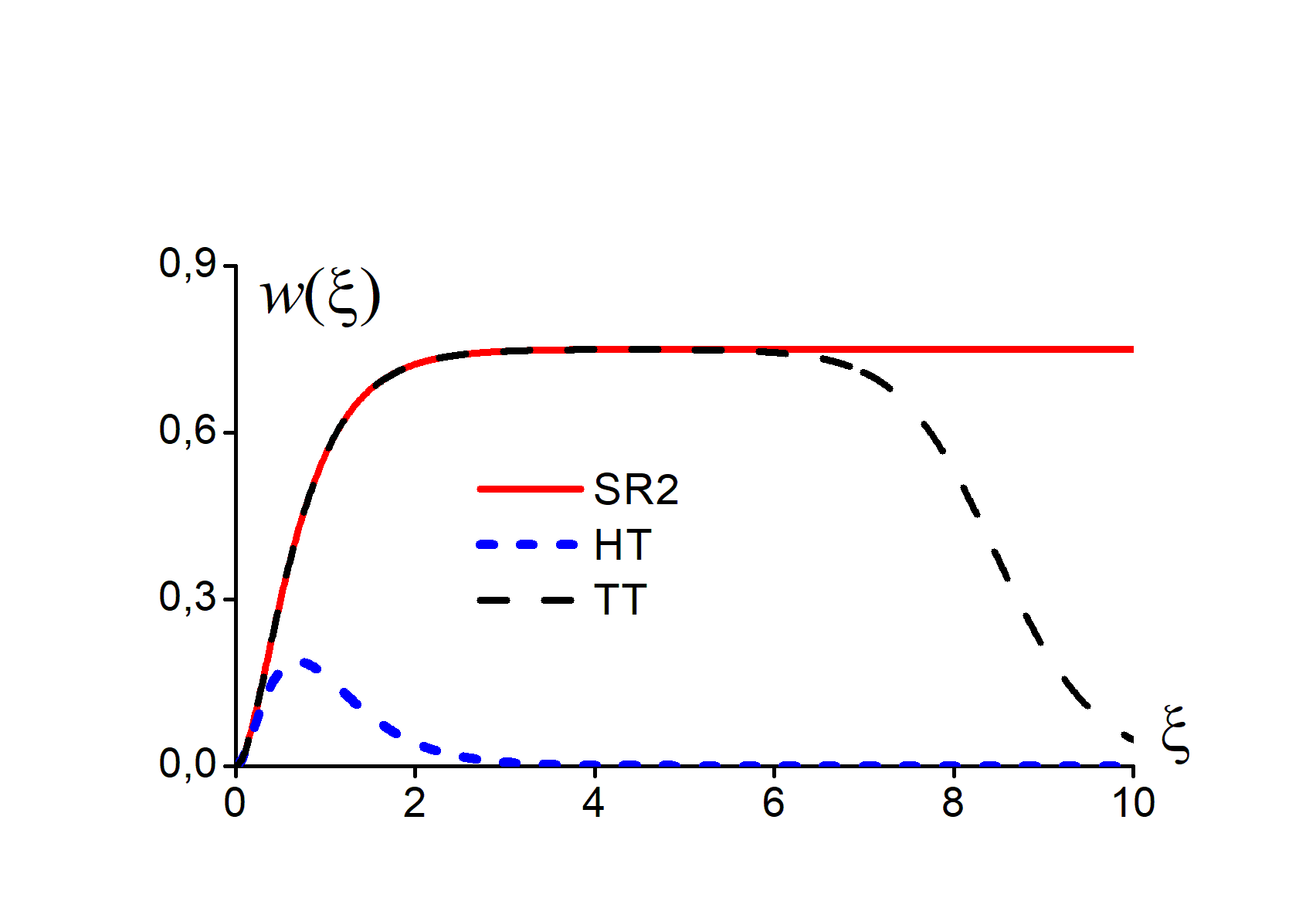}
 \caption{Potentials:   SR2  (\ref{SF-starobins}),  HT   (\ref{hilltop}) and TT (\ref{f(R)table-top}) with $p=10^7$.}
\label{fig1:potentials}
 \end{figure}

Consider scalaron  
potentials   in the form  (see Fig. \ref{fig1:potentials})\begin{equation}\label{W-main} 
W(\xi) = \mu^2 w(\xi)\,,
\end{equation}
where  $w$ does not depend on $\mu$. With this notation, the  scalaron potential of the SR2  model  \cite{Starobinsky:1980te}   is \footnote{Some difference in signs of (\ref{SF-starobins}), (\ref{hilltop}),(\ref{f(R)table-top}) compared to \cite{Starobinsky:1980te},\cite{Shtanov-Sahni2023} is a consequence of our metric signature.}: 
\begin{equation}\label{SF-starobins} 
 f(R)=R-\frac{R^2}{6\mu^2}\,,   \quad w(\xi) = \frac{3}{4} \left(1 - e^{-2\xi}\right)^2\,. \end{equation}

The HT   example \cite{Shtanov-Sahni2023}   deals with
\begin{equation}\label{hilltop}
    f(R)=\frac{R}{1+ {R}/(6\mu^2) },\quad w(\xi)=3 e^{-2\xi} \left(1-e^{-\xi}\right)^2\,.
\end{equation} 
The TT example  \cite{Shtanov-Sahni2023} uses two masses $m_1\ll m_2$    
\begin{equation}\label{f(R)table-top}
    f(R)=R\frac{1-R/m_1^2}{1+R/m_2^2}\,;   
\end{equation}
  the scalaron potential is
\begin{equation}\label{table-top+}
    w(\xi)=3 e^{-4\xi} (p+1) \left(\sqrt{p+e^{2\xi}}-\sqrt{p+1}\right)^2\,,\quad 
\end{equation}
where the scalaron mass is $\mu=m_1\sqrt{p}/\sqrt{6(p+1)}\approx m_1/\sqrt{6}$ and  $p= {m_2^2}/{m_1^2}$ will be considered fixed for different $\mu$. The larger $p$, the longer is the plateau in the graph of $w$ and the closer it is to the R2 model. 

In all cases (\ref{SF-starobins}),  (\ref{hilltop}) and  (\ref{f(R)table-top})) \footnote{The factors 3  and 6  corresponds to the definition of scalaron according to  (\ref{conformal_xi}) and ensure that   $\mu$ is the scalaron  mass.}  
\begin{equation}\label{w_xi_small}
\quad w(\xi)=3  \xi^2+O(\xi^3)\,,\quad w'(\xi)=6  \xi+O(\xi^2)\,,
\end{equation}
 $w(\xi)\ge 0$ and  it is sufficiently smooth. Below, in analytical estimates we will use (\ref{W-main}), (\ref{w_xi_small}); numerical calculations use  (\ref{SF-starobins}), (\ref{hilltop}), (\ref{f(R)table-top}).

 The potential $w(\xi)$ will be called  monotone if $ w'(\xi)\ge 0$,  $\forall \xi>0$. The  scalaron potential of the SR2  model  is  monotone, but (\ref{hilltop}) and (\ref{f(R)table-top}) are not.

Further, for definiteness, we  consider only positive branch of $\xi(r)$. This does not lead to a loss of generality in our case, because, as we shall see below, $\xi(r)\ne 0$.


\section{Reduction of equations}\label{section:reduction}
Denote $x=\mu r$ and use (\ref{W-main}); then a linear combination of (\ref{Ein_1-0a}),(\ref{Ein_2-0a}) yields   
 \begin{equation} \label{Ein_1-0}
\frac{d\chi}{dx} = 3x\left(\frac{d\xi}{dx}\right)^2 \,, 
\end{equation} 
where $\chi= {(\alpha+\beta)}/{2}$,
\begin{equation} 
\label{Ein_2-0}
\frac{d}{dx}\frac{\alpha-\beta}{2}  = -\frac{1}{x}+\frac{e^\beta}{x} \left[1-x^2w(\xi)\right]\,.
\end{equation} 
Equation (\ref{equation-xi-vacuum}) yields 
\begin{equation} 	
\label{equation-xi-of-x}
\frac{d}{dx}\left[x^2 e^{\frac{\alpha-\beta}{2}}\frac{d\xi}{dx}\right]= \frac{x^2}{6} e^{\chi} w'(\xi)   \,.
\end{equation}
 Note that, though system (\ref{Ein_1-0}), (\ref{Ein_2-0}), (\ref{equation-xi-of-x}) does not contain $\mu$, this parameter is implicitly present in the boundary conditions at $\infty$.  

In case of an  asymptotically flat SSS   configuration,  there must be  a weak-field region, where $|\xi|\ll 1$  and the metric takes on the Schwarzschild values
\begin{equation}
\exp(\alpha)=1- X_{g}/x=\exp(-\beta) \,, 
\label{Schwarzs-metric-x}  
\end{equation}
  $r_g=2M$,  $x\gg X_g\equiv \mu r_g\gg 1$; $M>0$ is the configuration mass.

Solution of equation (\ref{equation-xi-of-x}) in the weak field region has  exponentially growing and decaying modes (see Appendix \ref{Appendix_A}) and we choose the latter with asymptotics  (cf. \cite{Asanov1968, Asanov1974, Rowan1976})
\begin{equation}\label{xi_inf}
\xi(x)=  Q\left(\frac{X_g}{x}\right)^{1+ X_g/2}\exp{(-x)}\,,\quad  x\gg X_g\,,
\end{equation} 
here the constant $Q$ that characterizes the strength of the scalar field at infinity, will be called the “scalar charge”.  
The corrections to  (\ref{Schwarzs-metric-x})  due to input of the nonzero scalaron field (\ref{xi_inf}) in the weak-field regime can be found in Appendix \ref{Appendix_A}.

 We separate whole interval $(0,\infty)$ of the radial variable into a weak-field region 
$A=[r_0,\infty)$ with some appropriate $r_0$, and a ``scalarization region'' $B=(0,r_0)$, where there is a considerable deviation from the Schwarzschild metric and where one can hope to see a smoking gun of the modified gravity. 
We will limit ourselves to the case $r_0=(10^2\div 10^4) r_g$ and  $X_0\equiv \mu r_0\gg X_g$.

The value $r_0$, that separates regions A and B, can be related to the scalar charge $Q$ using (\ref{xi_inf}). Denote 
\begin{equation} 
   q=\frac{\ln Q}{X_g}   =\frac{r_0}{r_g}+\frac12 \ln{\left(\frac{r_0}{r_g} \right)}+\frac{\ln\xi_0}{X_g}+O\left(\frac{r_g}{r_0}\right)\,.
        \label{r0}
\end{equation}
For large $X_0$ the ``initial'' value $\xi_0 \lesssim 1/\sqrt{X_0}  $ (see Appendix \ref{Appendix_A}) can be varied   within several orders of magnitude and this has virtually no effect on $r_0$ defined by (\ref{r0}). 
Then \textbf{from Eq. (\ref{r0}) in the limit $M\mu\gg 1$}  we have 
 \begin{equation} 
   q= \frac{r_0}{r_g}+\frac12 \ln{\left(\frac{r_0}{r_g} \right)} \,.
        \label{r0+1}
\end{equation}

It is convenient  to introduce 
\begin{equation} \label{Y}
      Y=\frac{U}{X_0}\exp\left(\frac{\alpha-\beta}{2}\right)\,,\quad U=\frac{x}{X_0}\equiv \frac{r}{r_0}\,.
\end{equation}

   Multiplying (\ref{Ein_2-0}) by $\exp[(\alpha-\beta)/2]$, after simple transformations we obtain
  \begin{equation}
\label{Ein_2-ss}
\frac{dY}{dx}  =e^\chi \left[\frac{1}{X_0^2}- U^2w(\xi)\right]\,.
    \end{equation}
Using (\ref{Ein_1-0}), (\ref{equation-xi-of-x}) and   (\ref{Ein_2-ss}), one can  check by a direct differentiation 
\begin{equation}\label{dEds}
\frac{dI}{dx}= -\frac23 \frac{e^\chi}{X_0^2} 
\end{equation}
where
\begin{align}
I(x)=&Y\left[\left(x\frac{d\xi}{dx}\right)^2-1\right]+\frac{x}{3}e^\chi \left(\frac{1}{X_0^2}- U^2w(\xi)\right) \equiv\nonumber\\
&\equiv Y\left[\left(x\frac{d\xi}{dx}\right)^2-1\right]+\frac{x}{3} \frac{dY}{dx}\,.
\label{I(x)}
\end{align}
This relation will be used later to estimate $\zeta=\lim [xd\xi/dx]$ for $x\to 0$.

Having the approximate solution in  region A, one can extend it to region B. \textbf{To avoid exponentially large numbers in the transition A$ \to $B, we use a slightly different representation of the underlying system of equations than that used in \cite{ZhdStSht2024}.} 
We introduce a new variable $s: x\equiv X_0-s$ ($s\ge 0$) and reduce the equations (\ref{Ein_1-0}), (\ref{Ein_2-0}), (\ref{equation-xi-of-x}) to an equivalent  normal form ready for numerical integration. 
We have from (\ref{Ein_2-ss})
 \begin{equation}
\label{Ein_2-ss1}
\frac{dY}{ds}  =- e^\chi \left[\frac{1}{X_0^2}- U^2w(\xi)\right],\quad U(s)\equiv 1-\frac{s}{X_0}\,.
    \end{equation}
By denoting 
\begin{equation}\label{Z}
 Z={U^2} \exp\left(\frac{\alpha-\beta}{2}\right)\frac{d\xi}{ds}=X_0 U Y\frac{d\xi}{ds}\,,   
\end{equation}
we obtain from (\ref{equation-xi-of-x})  two first-order equations
\begin{equation}\label{dxids}
\frac{d\xi}{ds}=\frac{Z}{X_0UY} \,,
\end{equation}
\begin{equation}\label{dZds}
    \frac{dZ}{ds}= \frac16  U^2   e^{\chi} w'(\xi)   \,.
\end{equation}

On account of (\ref{dxids}),  equation (\ref{Ein_1-0}) yields 
 \begin{equation}   \label{Ein_1-sss}
\frac{d\chi}{ds}=-\frac{3 Z^2}{X_0 U Y^2}\,.    
\end{equation} 

\section{Transition between regions of weak and strong scalaron field  and initial conditions at $r_0$ }\label{section:transition}
To extend the  solution to region B by means of the system (\ref{Ein_2-ss}),(\ref{dxids}),(\ref{dZds}),(\ref{Ein_1-sss}), we need the initial conditions at $t= 0\, (r=r_0)$ inherent in the asymptotically flat configuration: 
\begin{equation}\label{metric_at_X0}
   e^{\alpha_0}=1- {r_g}/{r_0}=1- {X_g}/{X_0} =e^{-\beta_0}\,.
\end{equation}
 
 Equations (\ref{metric_at_X0}) must be supplemented by conditions for $\xi(t)$
\begin{equation}\label{field_at_X0}
   \xi(0)=\xi_0\,,\quad \xi'(0)= \xi_0e^{\beta_0/2}\,, 
\end{equation}
   where $\xi'\equiv d\xi/ds$ and we used (\ref{xiWKB_0_xi}). 
  
 In view of (\ref{restriction_xi0}),    in order for  the weak field approximation (\ref{Schwarzs-metric-x}),(\ref{xi_inf}) to be usable  for  $x=X_0$  ($s=0$), it is necessary to have $\xi_0\ll 1/\sqrt{X_0}$. We set 
\begin{equation}\label{Initial_xi}
\xi_0= (\sqrt{3}X_0)^{-1}\,,  
\end{equation}
 which corresponds to $dY/ds=-dY/dx= O(\xi^3)$ in (\ref{Ein_2-ss}) in view of  (\ref{w_xi_small}). 

Taking into account (\ref{field_at_X0}), 
\begin{equation}\label{initialZ_Y} 
    Y(0)=\left(1-\frac{r_g}{r_0}    \right)X_0^{-1}\,, Z(0)=\xi_0 \sqrt{ 1-\frac{r_g}{r_0} }\,.
\end{equation}

The transition from A to B near $r\approx r_0$ ($s=O(\sqrt{X_0}) $) is accompanied by rapid change of $e^\beta, Z, Y$.
Analytical and numerical estimates show that  in this region $e^\alpha$  and $U(t)\approx 1$  may be considered as slowly varied functions and can be eliminated from   (\ref{Y}),(\ref{Ein_2-ss1}). After combination with (\ref{xiWKB_0_xi}), which is valid for $|\xi|\ll 1$ these  considerations yield a simple  autonomous system of two first-order equations with respect to $\exp(-\beta/2)$ and $\xi$. Neglecting terms $\sim O(r_g/r_0)$ and $O(X_0^{-1})$ yields 
\begin{equation}\label{ebeta_est}
  e^{-\beta/2}  = 1+\frac32 X_0 (\xi^2-\xi_0^2) \, .  
\end{equation}
Using (\ref{xiWKB_0_xi}) and (\ref{ebeta_est}) allows us to infer  
\begin{equation}
\ln\frac{\xi}{\xi_0}+\frac34 X_0(\xi^2-\xi_0^2)-\frac{1}{2X_0}\ln^2\frac{\xi}{\xi_0}= s\,. 
\end{equation}
  In particular, for $\xi\gg\xi_0$
\begin{equation}\label{xi_appro}
 \xi \approx \sqrt{\xi_0^2+\frac{4s}{3X_0}}=\sqrt{\xi_0^2+\frac{4(r_0-r)}{3r_0}}\,. 
\end{equation}
 and 
\begin{equation}\label{beta_appro}
     \exp(-\beta/2)\approx 1+2 s=1+2X_0 \frac{ (r_0-r)}{ r_0} \,.
 \end{equation}
 Surprisingly, numerical simulations show that (\ref{xi_appro}) and (\ref{beta_appro}) can be extended for larger $\xi$, namely,  for $0<s\lesssim 0.1X_0$ ($0.9\lesssim r/r_0<1$). Also, we have a good fit of $\alpha$ for almost all region B
 \begin{equation}\label{alpha_B}
     \alpha(s)\approx 2\ln{[(X_0-s)/X_0]}=2\ln{(r/r_0)}\,.
 \end{equation}
 
\section{Numerical solutions in the scalarization region }\label{section:numerical}

 \begin{figure}[h!]  	\centering
\includegraphics[width=0.45 \textwidth]{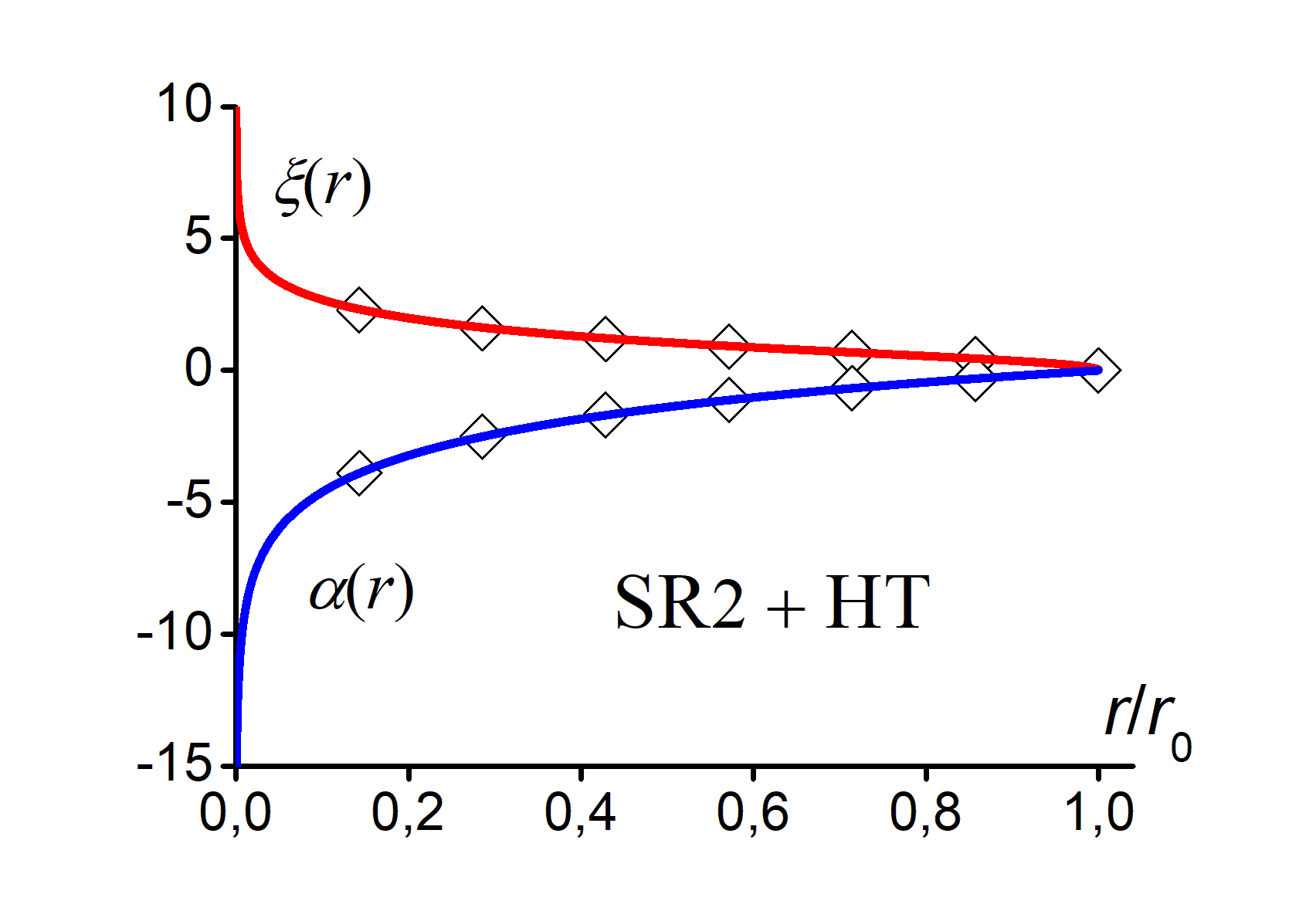}
\caption{Universal behavior of $\alpha(r)$ and $\xi(r)$ in case of   potentials (\ref{SF-starobins}),   (\ref{hilltop}) and (\ref{f(R)table-top}) regardless of $M\mu=10\div 10^{20}$ and  various   $r_0=(100\div 10^{10})r_g$. \textbf{In case of the TT model (\ref{f(R)table-top}), which is intermediate between SR2 and HT,} we used  $p=1,10,1000$.  All the curves are  practically the same after rescaling to $r/r_0$; e.g., rhombs illustrate  the coincidence of the SR2 model (\ref{SF-starobins}) and  the HT case  (\ref{hilltop}) with the same parameters.  Outside the scalarization   regions ($r\ge r_0$)   $\alpha\approx -r_g/r$ and $\xi$ is given by ($\ref{xi_inf}$).}
 	\label{fig2:alpha(r)}
 \end{figure}
 
 \begin{figure}[h!]  	\centering
\includegraphics[width=0.35 \textwidth]{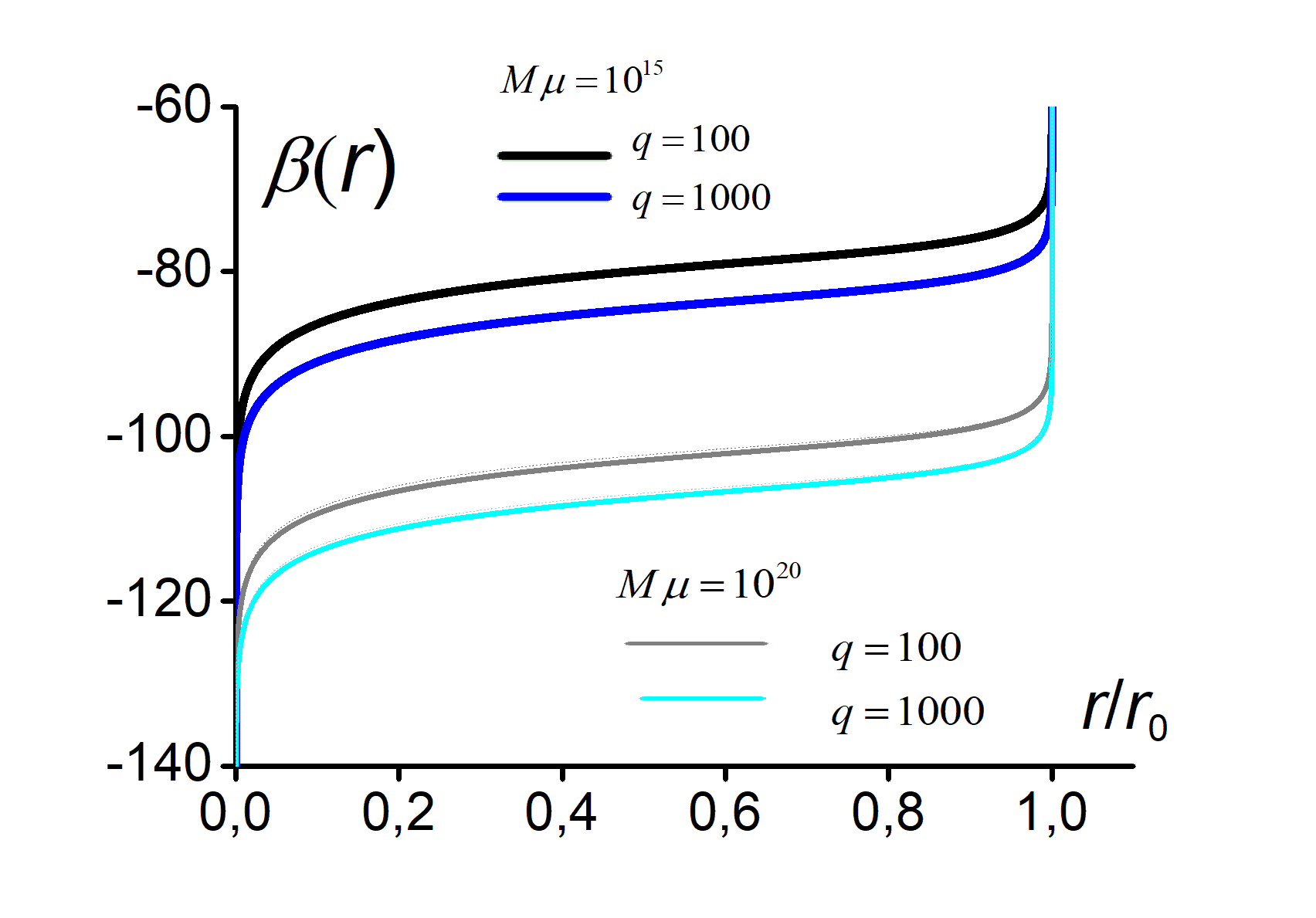}
 	\caption{\textbf{The figure shows how $\beta(r)$ depends on the parameters in case of the SR2  model.}  Two upper curves show the case $\mu M=10^{15}$, for $q=100$ (black) and $q=1000$ (blue). Two lower ones:   $\mu M=10^{20}$,    $q=100$ (gray) and $q=1000$ (light blue). The curves are cut off at the top right corner, where for all the cases $\beta(r_0)\approx r_g/r_0$. Outside the scalarization   region ($r\ge r_0$)   $\beta(r)\approx r_g/r\ll 1$. \textbf{The HT and TT models also show similar dependence on the parameters.}}
 	\label{fig3:beta(r)}
 \end{figure}
\begin{figure}[h!]  	\centering
\includegraphics[width=0.5 \textwidth]{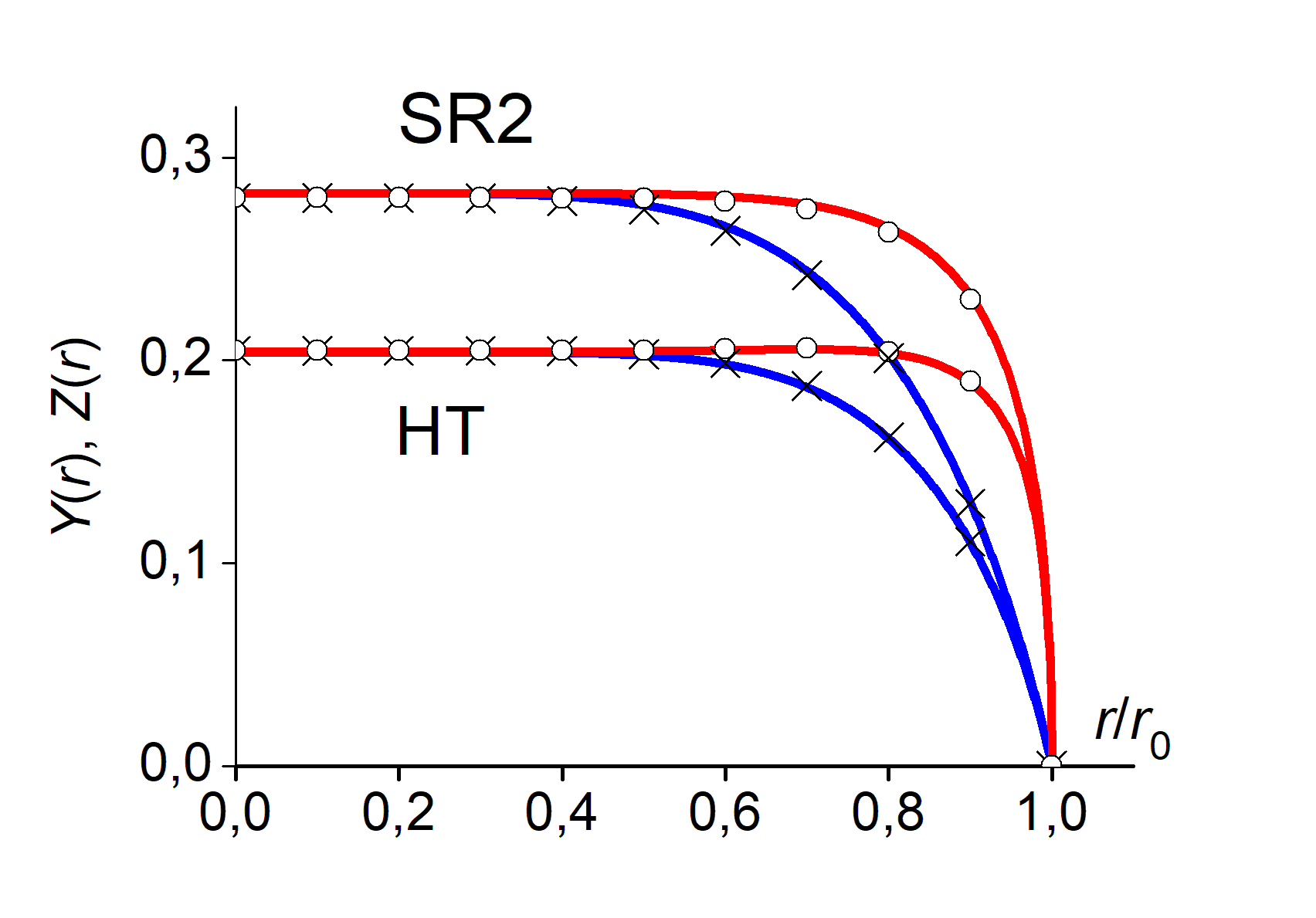} 	\caption{Dependencies $Y(r)$ (blue) and $Z(r)$ (red) in the scalarization regions after rescaling to $r/r_0$ for the $R2$  model (\ref{SF-starobins}) (two upper curves)  and for HT scalaron  potential (\ref{hilltop}) (lower curves); $M\mu= 10^{20}$, the size of the scalarization region is $r_0=10^2\div10^{10}r_g$. The small circles on $Z$ curve and crosses on $Y$ curve show the same dependencies for $M\mu=10$. \textbf{But the difference between models is considerable.  All the curves have a smooth continuation for $r>r_0$ (not shown here).  }}
\label{fig4:yz}
 \end{figure}
\begin{figure}[h!]  	\centering
\includegraphics[width=0.5 \textwidth]{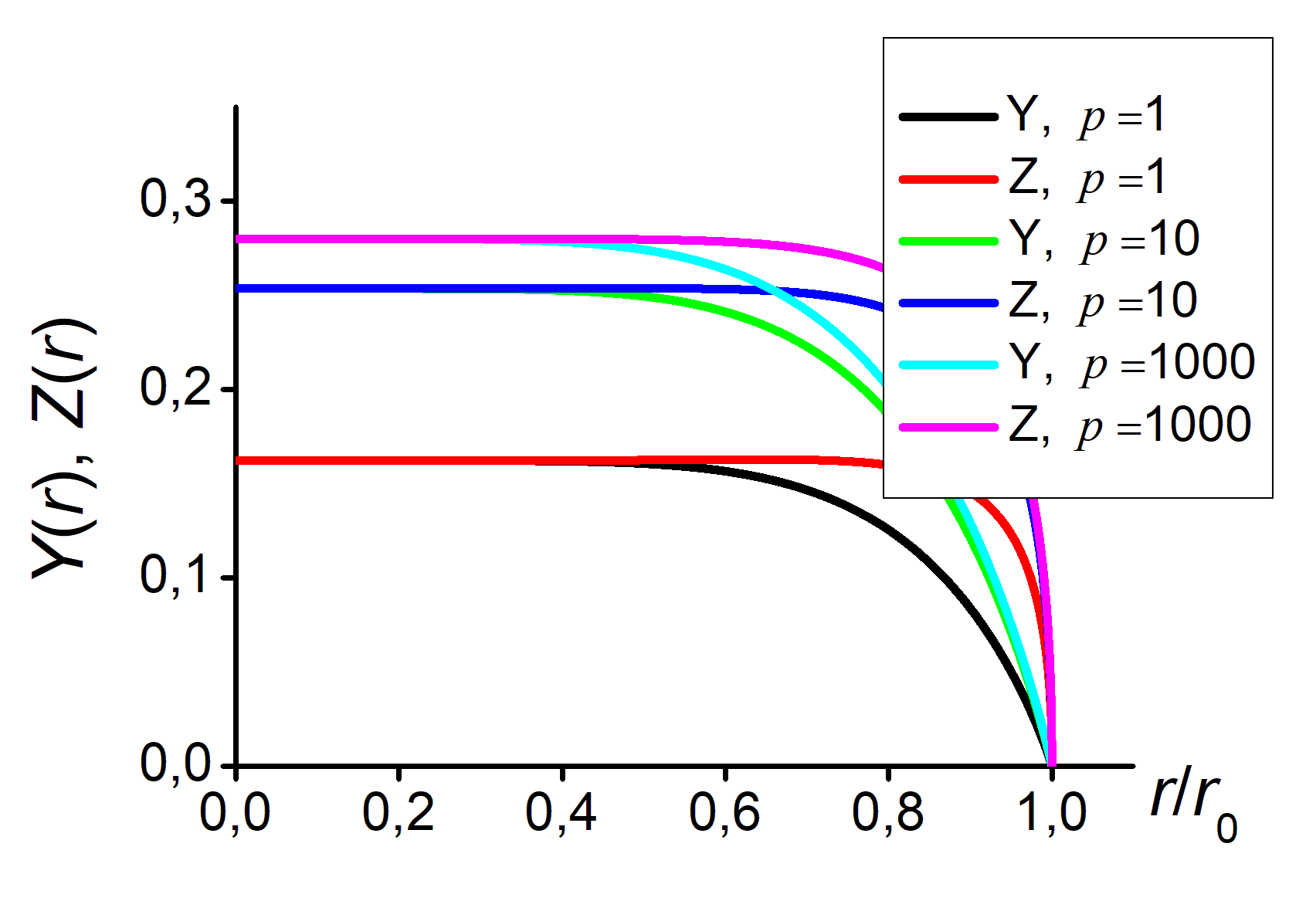}
 \caption{\textbf{The figure shows differences between TT models (\ref{f(R)table-top}) with different $p=1,10, 1000$ ($M\mu=10^{15},\, q=1000$) for $Y(r)<Z(r)$   in the scalarization regions after rescaling to $r/r_0$. At the same time, the curves with fixed $p$ are practically independent of the parameters from interval $M\mu=10\div 10^{20},\, q=100\div 10^5$.} }
\label{fig5:YZTt-1-10-1000}
 \end{figure}
\begin{figure}[h!]  	\centering
\includegraphics[width=0.45 \textwidth]{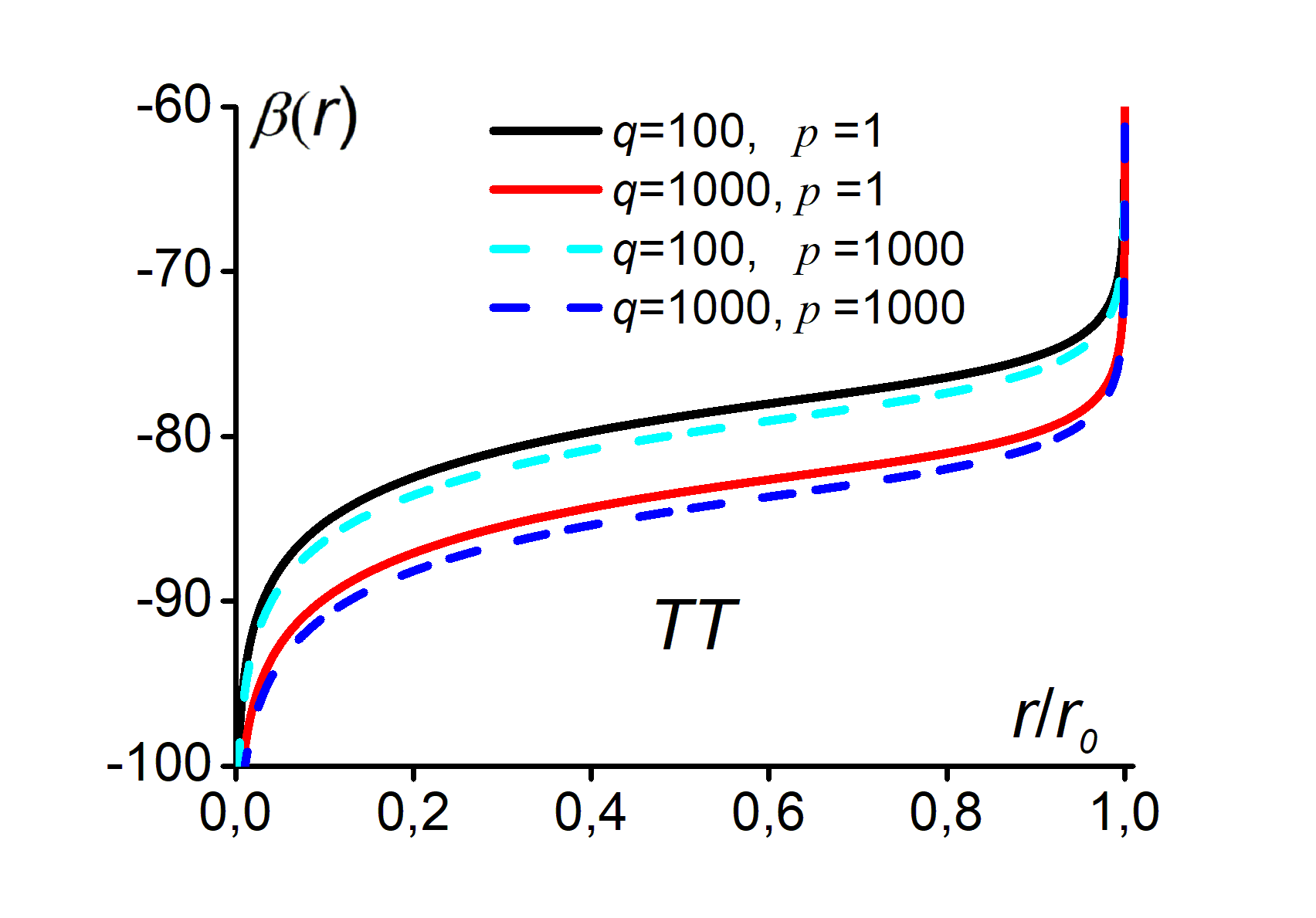}
 \caption{\textbf{There is an obvious difference of $\beta(r)$ for the TT models (\ref{f(R)table-top}) with different $p=1$ (solid) and $p=1000$ (dashed).} This figure shows $\beta(r)$ in the scalarization region after rescaling to $r/r_0$; in both cases $M\mu=10^{15}$, the sizes of the scalarization regions are $r_0=100 r_g$   and $r_0=1000r_g$. The behavior at the top right corner is analogous to Fig. \ref{fig3:beta(r)}; outside the scalarization   region $\beta(r)\approx r_g/r\ll 1$.}
\label{fig6:beta-tt}
 \end{figure}

 We  solved numerically system (\ref{Ein_2-ss}),(\ref{dxids}),(\ref{dZds}),(\ref{Ein_1-sss})  in case of potentials (\ref{SF-starobins}),   (\ref{hilltop}) and (\ref{f(R)table-top}) for $s\in [0, X_0)$. The initial conditions for integration where chosen according to (\ref{metric_at_X0}), (\ref{field_at_X0}), (\ref{Initial_xi}).   The regions of parameters  were as follows: $M\mu=1\div 10^{20}$, $\log(q)=2\div {10}$. In case of (\ref{f(R)table-top}) we considered $p=1\div1000$. 
The results of the numerical studies are shown in Figs. \ref{fig2:alpha(r)}--\ref{fig6:beta-tt}.
The graphs $\alpha(r)$, $\xi(r)$   are shown in  
Fig. \ref{fig2:alpha(r)}; they turn out to be practically the same for all $\log(M\mu)>5$. 

Though the size of the scalarization region depends on $q$ according to  (\ref{r0+1}), dependencies $\alpha$ and $\xi$  after rescaling $r\to r/r_0$ are almost identical even for different models: the  r.m.s. deviation between is of the order of $10^{-2}\div 10^{-3}$. However, the graphs of $Y(r)$, $Z(r)$   and especially $\beta(r)$ differ  for different models significantly. \textbf{ The differences between $\alpha(r)$ for different models decrease to zero with increasing $qM\mu$  as  shown in Figs. \ref{fig1ab:differences}, \ref{fig1ab:differencesSt-TT} (see Appendix \ref{Appendix_B}). The differences between $\xi(r)$ for SR2 and HT models turn out to be small but do not decrease to zero with increasing $qM\mu$; however, this is the case for SR2 and TT models as $p\to \infty$ (i.e., when the `shelf' in the scalaron potential is large).} 
At the same time, the `universality' of $\alpha$ and $\xi$  does not mean that whole solutions for different $M\mu$ are identical, which is clearly visible in  Figs. \ref{fig3:beta(r)}, \ref{fig4:yz}, \ref{fig5:YZTt-1-10-1000}, \ref{fig6:beta-tt}) for other components of the solutions.

In the region B, functions $\xi(x)$, $Y(x)>0$ and $Z(x)>0$ do not change their signs   $\forall x\in (0, \infty)$.  
 In a non-trivial case ($Q\ne 0$)   the graphs of $Y(r)$ and $Z(r)$ reach a plateau  at   $0\le r\lesssim 0.5 r_0$ and 
 there  exist finite limits 
\begin{equation} \label{limits_rto0}
Y_C=\lim\limits_{x\to 0} Y(x)>0\,,\quad  Z_C=\lim\limits_{x\to 0}Z(x)>0\,\,.
\end{equation} 
The existence of  limits (\ref{limits_rto0}) is ensured by the rapid decrease of $e^\chi\sim e^\beta$  as $x$ decreases (see Figs. \ref{fig3:beta(r)}, \ref{fig6:beta-tt}; for $x\to 0$ functions $\xi(x), \alpha(x), \beta(x)$ have a logarithmic behavior. This is in agreement with general results concerning SSS configurations of General Relativity  with scalar fields \cite{ZhdSt} and configurations of the R2 Starobinsky $f(R)$ gravity \cite{ZhdStSht2024} dealing with monotone potentials. 

In case of non-monotone TT and HT  potentials, the results of 
\cite{ZhdSt} and \cite{ZhdStSht2024} cannot be directly applied.  In this case   the potential $ w(\xi)\ge 0$ is non-decreasing at least for some  $\xi\le \xi_m$, where $ \xi_m>1$ is independent on $\mu$ and it decreases to zero for $\xi>\xi_m$, but not very fast. However, the qualitative behavior of solutions is completely similar to the case of the SR2 model; moreover, the dependences $\alpha$ and $\xi$ against $r/r_0$ are almost the same (Figs. \ref{fig2:alpha(r)}). The condition $\xi\ne 0$ is also valid for non-monotonic potentials (\ref{hilltop}), (\ref{f(R)table-top}), see Figs.  \ref{fig2:alpha(r)}, \ref{fig4:yz}, \ref{fig5:YZTt-1-10-1000}. The reason is that the main changes in $Z(r)$ and $Y(r)$ occur  in the region where $w(\xi)$ is monotonous;  outside  this region (for $\xi>\xi_m$) variations of $Z(r)$ and $Y(r)$ are damped by exponentially small factor $\exp(\chi)$ in the right-hand sides of (\ref{Ein_2-ss1}) and (\ref{dZds}).


The specificity of our case is that for large enough  $M\mu$ and $q$ we can estimate parameter
\begin{equation}\label{LimZeta}
\zeta =  \ \frac{Z_C}{Y_C} =-\lim\limits_{x\to 0}
\left(x\frac{d\xi}{dx}\right)=1 \,,         
\end{equation}
as can be seen from Figs. \ref{fig4:yz}, \ref{fig5:YZTt-1-10-1000}  for all the models  involved. 

This can be explained using (\ref{I(x)}). For $x\to 0$ 
\begin{equation*}
I(0)= Y_C\lim\limits_{s\to 0}\left[\left(x\frac{d\xi}{dx}\right)^2-1\right]   
\end{equation*}
Note that $I(X_0)\sim O(X_0^{-2})$, whereas $Y(x)\sim O(1)$ independently on $X_0$.
 Using $\chi(x)\le \chi(X_0)\le 0$ in virtue of (\ref{Ein_1-sss}),  we  have a rough estimate
\begin{align*}  
Y_C\lim\limits_{x\to 0}&\left[\left(x\frac{d\xi}{dx}\right)^2-1\right]=\\
    & =I(X_0)+\frac23\int\limits_0^{X_0} \frac{e^\chi(s)}{X_0^2} \lesssim O(X_0^{-1})\,,
\end{align*}
whence 
\[
\lim\limits_{x\to 0} \left(x\frac{d\xi}{dx}\right)^2 = 1+O(X_0^{-1})\,.
\]
Taking into account the sign of $d\xi/dx$ in case of asymptotic  flatness,  we have (\ref{LimZeta}).

The   leading terms of the asymptotics near  the center of the configuration ($r\to 0$) for all potentials: 
\begin{equation} 
\xi(r)\sim -  \ln(r/r_g)\,,\quad  \chi(r)\sim 3 \ln(r/r_g)\,,
\end{equation}
\begin{equation} 
\beta(r)\sim 4\ln(r/r_0)\,
\label{asymt-center}
\end{equation}
and in case of $\alpha$ approximate formula (\ref{alpha_B}) works for $0<r/r_m<1$.

\section{Jordan frame}\label{Jordan_frame}
Above results are obtained in the Einstein frames appropriate to each model with metric $\hat g_{\mu\nu}$. \textbf{ The physical metric $g_{\mu\nu}$  of the Jordan frame  is connected with it according to (\ref{conformal_xi}), (\ref{metric}):} 
\begin{align*} 
& g_{00}=e^{\alpha-2\xi}\,,\quad  g_{11}= e^{\beta-2\xi}\,,   \\ & g_{22}=-r^2e^{-2\xi}\,,\quad  g_{33}=-r^2e^{-2\xi}\sin^2(\theta)\,.
\end{align*}

\textbf{Transition to the curvature  (Schwarzschild) coordinates yields} 
\begin{equation} \label{metric_Jordan} 
ds_J^2 =g_{\mu\nu}dx^\mu dx^\nu =A_J(\rho)  dt^{2} -B_J(\rho)d\rho^{2} -\rho^{2} dO^2 ,
\end{equation}
where 
\begin{align}\label{alpha_Jordan}
      A_J(\rho)=e^{\alpha-2\xi}\,,\quad B_J(\rho)= e^{\beta-2\xi}\left(\frac{dr}{d\rho}\right)^2 \,,
\end{align}
\[
\alpha\equiv \alpha(r),\,\beta\equiv\beta(r), \, \xi\equiv \xi(r),\, 
 \rho\equiv r\exp[-\xi(r)]\,.
\]
\textbf{Using (\ref{dxids}) that can be rewritten as $xd\xi/dx=-Z/Y$, we get }
\begin{equation}\label{beta_Jordan}
   \ln B_J(\rho)\equiv \beta(r)-2\ln\left(1+\frac{Z(r)}{Y(r)}\right)  
\end{equation}

\textbf{With the help of (\ref{LimZeta}) one can estimate  leading terms of asymptotic formulas near the center. Directly from  (\ref{Ein_1-0}), (\ref{Ein_2-0}), (\ref{dxids}) we get}  
\begin{equation} \label{asympt_r}
x\frac{d\alpha }{dx} \sim 3\zeta^2-1\,, \quad x\frac{d\beta }{dx}\sim 3\zeta^2+1\,,\quad x\frac{d\xi}{dx} \sim -\zeta\,. 
\end{equation}
For $\rho\to 0$ we have $\ln \rho\sim (1+\zeta) \ln r$ and\footnote{Here we corrected an error in formula (48) for $a$ of \cite{ZhdStSht2024}.}
\begin{equation}\label{asympt_Jordan}
 \ln A_J(\rho)\sim 
a \ln(\rho/r_g)\,,\quad \ln B_J(\rho)\sim  b \ln(\rho/r_g)\,,
\end{equation}
where
\begin{equation*}
a =\frac{3\zeta^2+2\zeta-1}{1+\zeta}=2\,, \quad  b=\frac{3\zeta^2+1}{1+\zeta}=2\,.
\end{equation*}
The qualitative behavior of metric components in the Jordan frame looks similar to the Einstein frame. As can be seen from  (\ref{alpha_Jordan}), $\ln A(\rho)$ is independent on $M\mu$ for all models, like $\alpha$ and $\xi$. The behavior of $\ln B$ looks similar to Figs. \ref{fig3:beta(r)}, \ref{fig6:beta-tt}. 

Standard investigations of the test-body motion in  spherically symmetric space-time (\ref{metric_Jordan}) lead to the one-body problem with effective potential
\begin{equation}\label{VCO}
    V_{TB}(r)=A_J\left(1+\frac{L^2}{\rho^2}\right)=e^{\alpha-2\xi}+ \frac{e^\alpha}{r^2}L^2 \,,
\end{equation}
where $L=\rho^2 d\varphi/ds_J$. 

The existence and stability of the test body circular orbits  around the SSS configuration is determined by occurrence of minima of (\ref{VCO}). 

In view of (\ref{alpha_B}) $ {e^\alpha}/{r^2}\approx const $; also  $\alpha-2\xi$ is monotonically increasing as a function of $r$ for all the models (see Fig \ref{fig2:alpha(r)}). Then $V_{TB}(r)$ is monotonically increasing (at least, for moderate values of $L^2$); this means either the absence of circular orbits in the scalarization region or presence of weakly stable orbits. The latter is possible because  the relation (\ref{alpha_B}) is approximate and  small deviations of ${e^\alpha}/{r^2}$ from constant may lead to an appearance of weakly stable circular orbits for large $L^2$. Indeed, detailed  investigations show that stable circular orbits with very large values of $L^2$ can really exist near the center; however, the corresponding minima of $V_{TB}(r)$ are very shallow and test particles will not be able to remain in these orbits for long in the presence of even small non-gravitational forces.

\section{Discussion}
\label{section:discussion}
\textbf{ We studied the  SSS solutions  in different $f(R)$ models (\ref{SF-starobins}), (\ref{hilltop}), (\ref{f(R)table-top}) under the condition of asymptotic flatness. It is known from general theory \cite{ZhdSt} that such solutions are regular  for all radii $r>0$ except the naked singularity at the center. This is true in the coordinates, corresponding to metric (\ref{metric}) in the Einstein frame and therefore for $\rho$ variable  (\ref{metric_Jordan}) in the Jordan frame.  These solutions represent putative physical objects with the properties described below.\footnote{At the same time, these results may be regarded as some exterior solutions that must be continuously matched to the internal solutions that involve, e.g., hydrodynamical energy-momentum tensor}. The solutions are well defined for all $M>0$, $\mu>0$ and $Q\ne 0$ \cite{ZhdSt}. However, we focus on large  $M\mu$ and $q$. The latter means that the  strong field (scalarization) region B  with radius  $r_0\sim q r_g$ must be sufficiently large, as well as the value of the scalar charge $Q$.}\footnote{It should be noted that in the opposite case $Q\to 0$ the solutions behave differently \cite{ZhdStSht2024}. In this case, the weak-field region A, where  $e^\beta\approx (1-r_g/r)^{-1}$, will come very close to the value $r=r_g$, which will lead to a strong peak near this point.} 

\textbf{For large $M\mu$ and $q$, we found  analytic approximate solutions in the weak-field region (A) and in the small transition interval between A and B. 
It is the requirement of large $M\mu$ and $q$ leads to specific properties concerning the similarity of solutions. On the other hand, this requirement is  necessary in our treatment, because it has been used in the approximations resulting in the  choice $X_0$ and initial conditions at this point, which are used to extend the solutions for $x<X_0$. In our simulations we typically used $q\gtrsim 100$, $M\mu\sim 100\div 10^{20}$.}

In region B $\alpha, \xi, Y, Z$ as functions of $r/r_0$ tend to limiting  dependencies  (separately for each model) for sufficiently large $M\mu$. These properties  do not follow from simple scaling arguments for the entire solution:  as is clearly seen from Figs. \ref{fig3:beta(r)} and  \ref{fig6:beta-tt}, $\beta(x)$ varies significantly for different $M\mu$. The reason is that although the system (\ref{Ein_1-0}), (\ref{Ein_2-0}), (\ref{equation-xi-of-x}) does not depend on $\mu$, the initial conditions at $r_0$ do,  affecting some components of the solutions. This has been tested numerically for   $M\mu=10\div 10^{20}$ and $r_0/r_g\sim 10^2\div 10^{10}$. 

Moreover, the $\alpha, \xi$ plots (Fig. \ref{fig2:alpha(r)}) are universal for all these models (\ref{SF-starobins}), (\ref{hilltop}), (\ref{f(R)table-top}), although the scalar potential (\ref{SF-starobins}) is monotone, while (\ref{hilltop}), (\ref{f(R)table-top}) deal with non-monotone ones and have different asymptotics for $\xi\to \infty$. 
In all these cases we obtained  parameter $\zeta\approx 1$ (\ref{LimZeta}), which  determines asymptotic properties of the solutions near the center $r=0$.

\textbf{Comparison of models shows that the TT case is intermediate between SR2 and HT scalaron potentials. The larger the horizontal shelf in the TT potential (see Fig. \ref{fig1:potentials}), the closer are $\alpha(x)$ and $\xi(x)$  to the corresponding components of the SR2 solution. The other key reason of the `universality' is the same behavior of the reduced potential $w(\xi)$ near the  minimum. It turns out that in case of large $M\mu$, the behavior of $w(\xi)$ even for moderate $\xi$ is not essential: the  contribution of $w(\xi)$ and $w'(\xi)$ near the origin is suppressed by small factor $\exp(\chi(x))\to 0$ as $x\to 0$ in Eqs. (\ref{equation-xi-vacuum}) and (\ref{Ein_2-ss}) (cf.  analogous estimates in \cite{ZhdSt}). This property is expected for much wider class of potentials; e.g., we confirmed this for potentials with power-law growth at large  $\xi$.}

Our results have been obtained in the Einstein frame. The metric coefficients in the original Jordan frame (physical metric) are given by  (\ref{metric_Jordan}), (\ref{alpha_Jordan}), (\ref{beta_Jordan}). In the Jordan frame, distributions of stable circular orbits outside the scalarization region are the same as in case of the Schwarzschild metric. Inside this region, we found that the weakly stable (that is, technically unstable) circular geodesics for this metric may occupy a central area surrounded by a ring, where there are no circular orbits at all. It can be concluded that, if such a $f(R)$ configuration really exists, the size of the central empty region of an accretion disk could be either significantly larger than that of a typical black hole or, if $r_0$ is large enough, the accretion disk may be practically absent. This could be an observational signature to distinguish these configurations from normal black holes.

The universality of some elements of the SSS solutions can be used to extend earlier results \cite{ZhdStSht2024} regarding radial stability to large values of $M\mu$. This is discussed in Appendix \ref{SS_perturbations}.  
These considerations can be applied to the SR2, HT  and TT   models. The results of Appendix \ref{SS_perturbations} suggest that dark objects (distinct from ordinary black holes) without accretion disks may exist; however, further stability studies are needed to answer this question definitively.

\vskip3mm \textit{Acknowledgements.} I am grateful to Yu. Shtanov and O. Stashko for useful discussions.  
This work was  supported by the National Research Foundation of Ukraine under project No. 2023.03/0149.

\appendix
\section{Approximate solutions in the weak field region} \label{Appendix_A}
Consider (\ref{equation-xi-of-x}) for $0<\xi\ll 1$ assuming zero field at the infinity and $x\equiv r\mu\gg X_g\equiv \mu r_g$.  After linearizing of the right-hand side of (\ref{equation-xi-of-x}), on account of (\ref{w_xi_small}), we have the equation for a linear scalar field $\xi$
\begin{equation} 	
\label{equation-xi-linear}
\frac{d}{dx}\left[x^2 e^{\frac{\alpha-\beta}{2}}\frac{d\xi}{dx}\right]=   {x^2}  e^{\frac{\alpha+\beta}{2}} \xi   \,.
\end{equation}
Here we rely on the results of \cite{ZhdStSht2024} (see Appendix D). 
 Substitution  $\xi(x)=\exp[ u(x)]$  yields 
\begin{equation}\label{xiWKB_exact}
\left(\frac{du}{dx}\right)^2+ \frac{d^2u}{dx^2}+ \left[\frac{d}{dx}\left(2\ln x+\frac{\alpha-\beta}{2}\right)\right]=e^\beta     
\end{equation}
 The main contribution $u\approx u_0$ for $x\gg X_g$ and moderate $\alpha,\beta$ is due to the first term on the left-hand side of (\ref{xiWKB_exact})
 \begin{equation} 
\left(\frac{du_0}{dx}\right)^2=
e^{\beta}\left[1+O\left( {X_g^{-1}}\right)\right]   \,. \label{WBK_zero}
\end{equation}
There are exponentially growing and decaying modes  and we choose the latter:
\begin{equation}
\label{xiWKB_0}
\frac{du_0}{dx}=-e^{\beta/2}\left[1+O\left( {X_g^{-1}}\right)\right]    \,.
\end{equation}  
Integration of this equation in case of Schwarzschild metric   (\ref{Schwarzs-metric-x}) yields  
\begin{align}
&u_0(x)=-x\sqrt{1-\frac{X_g}{x}}-X_g\ln\left[1+\sqrt{1-\frac{X_g}{x}}\right]-\nonumber\\&\frac{X_g}{2}  \ln\left(\frac{x}{X_g}\right)+C_1 \approx -x-\frac{X_g}{2}  \ln\left(\frac{X_g}{x}\right)+C_2 \,, \label{A5}
\end{align}
$C_1,C_2=const$. 
 The next order correction $u=u_0+u_1$  in (\ref{xiWKB_exact})  leads to
\begin{equation}\label{xiWKB_1}
u_1(x)=- \ln x -\alpha/4+C_3 \,,  \end{equation} 
$C_3=const$.
For sufficiently  large $\mu$ we have $x\gg |\ln x+\alpha/4|$, which justifies the approximations. 
For $x\gg X_g$ from (\ref{A5}),(\ref{xiWKB_1})  we get (\ref{xi_inf}). 

Now we must estimate  corrections to $\alpha(x)$ and $\beta(x)$ using (\ref{xiWKB_0}) in the region where (\ref{Schwarzs-metric-x}) can be used.

 According to  (\ref{Schwarzs-metric-x}) $\chi(\infty)=0$, then the correction is
\begin{equation*}
\chi(x)=-3\int\limits_x^\infty dx' x'\left(\frac{d\xi}{dx}\right)^2\,.
\end{equation*}
From  (\ref{xiWKB_0}) we have
\begin{equation}
\label{xiWKB_0_xi}
\frac{d\xi}{dx}\approx -e^{\beta/2}\xi   \,.
\end{equation} 
then using   
(\ref{Ein_1-0}) and (\ref{Schwarzs-metric-x})
\begin{equation}
  \frac{d\chi}{dx} = 3xe^\beta\xi^2\approx 3x \xi^2 \,.  
\end{equation}

Taking into account   (\ref{xi_inf}), using integration by parts, we get
\begin{equation}\label{appro_chi}
   \chi(r)= -\frac32 x  \xi^2(x)\left[1+O\left(\frac{X_g}{x}\right)\right]\,.
\end{equation}
Equation (\ref{Ein_1-0a}) can be rewritten as 
\begin{equation*}
 \frac{d}{dx}\left[x( e^{\frac{\alpha-\beta}{2}}-1)\right]  = e^\chi-1-x^2 e^\chi w(\xi) \,.
\end{equation*}
Taking into account $\chi(\infty)=0$
\begin{equation*}
  x( e^{\frac{\alpha-\beta}{2}}-1) \approx -r_g- \int\limits_x^\infty \left\{e^\chi-1-x^2 w(\xi)\right\} \,,
\end{equation*}
whence using (\ref{w_xi_small}) and (\ref{xi_inf}) in the right-hand side, we obtain a leading contribution 
\begin{equation}\label{appro_alpha-beta}
   e^{\frac{\alpha-\beta}{2}}=1-\frac{X_g}{x}+  \frac32 x \xi^2(x)\left[1+O\left(\frac{X_g}{x}\right)\right] \,,
\end{equation}
Therefore, in order that (\ref{Schwarzs-metric-x}) can be used at $x=X_0$ with a sufficient accuracy, one must have 
\begin{equation}\label{restriction_xi0}
 X_0\xi_0^2  \ll 1\,.
\end{equation}
 This shows that one cannot choose $X_0$ sufficiently large without changing $\xi_0$. 


\section{Comparison of models}
 \label{Appendix_B}
Differences between models are not visible in Fig. \ref{fig2:alpha(r)}. In this connection deviations from the `universality' are assessed  here in more detail for various $q$ and $M\mu$. It turns out that the deviations can be characterized by one parameter $qM\mu$.  The differences between TT and SR2 tends to practically zero as $p$ increases. 
 \begin{figure}[h!] \centering
 	\includegraphics[width=0.35 \textwidth]{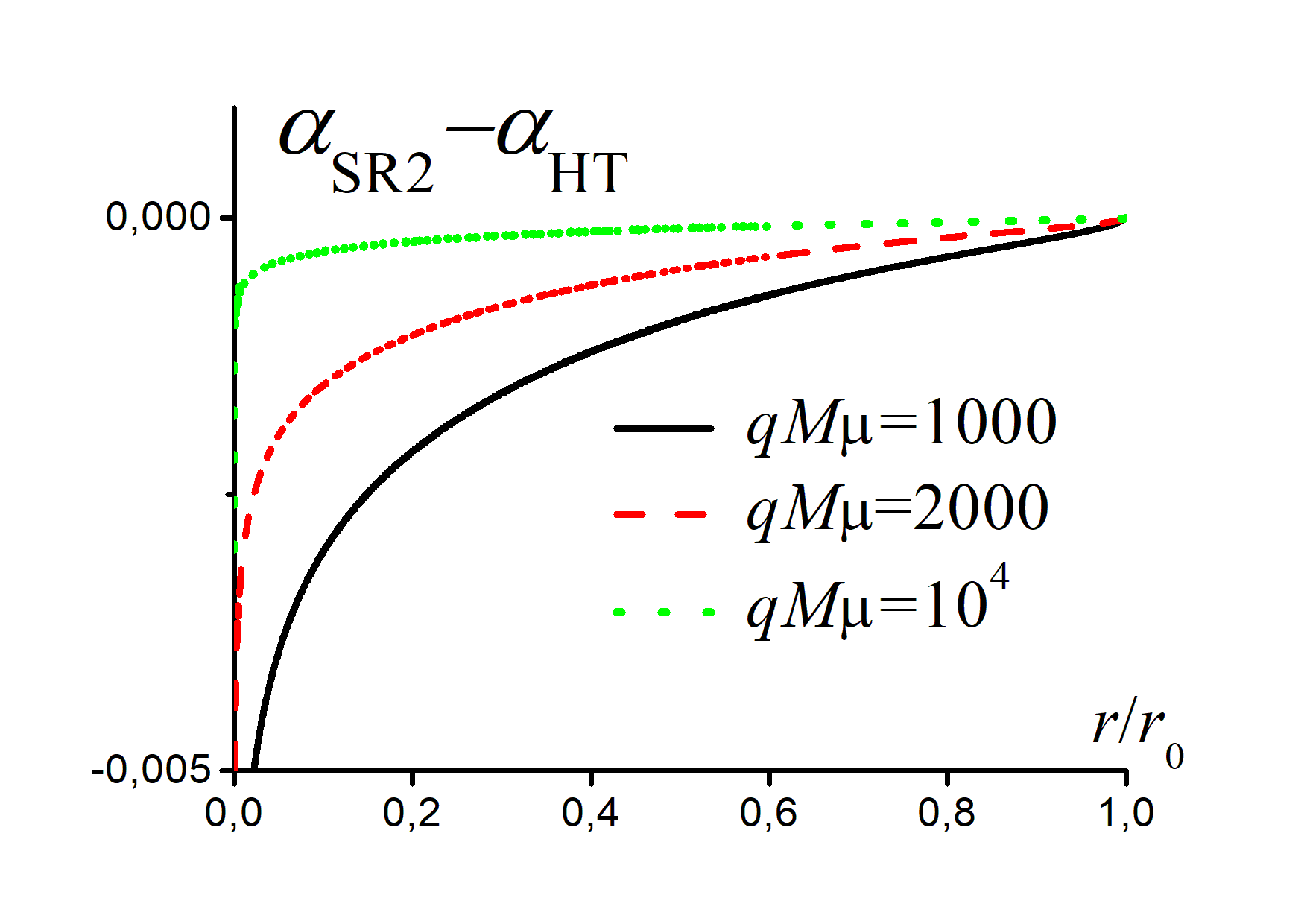}
 	\caption{Difference of $\alpha(r)$ for  SR2 and HT models with the same parameters $q$, $M\mu$ in the scalarization region, where $q$ is defined according to (\ref{r0+1}). The difference tends to zero as $qM\mu$ increases; this illustrates an  universality of the limit $\alpha$ as $qM\mu\to infty$.}
 	\label{fig1ab:differences}
 \end{figure}
 \begin{figure} [h!]  	\centering
 	\includegraphics[width=0.35 \textwidth]{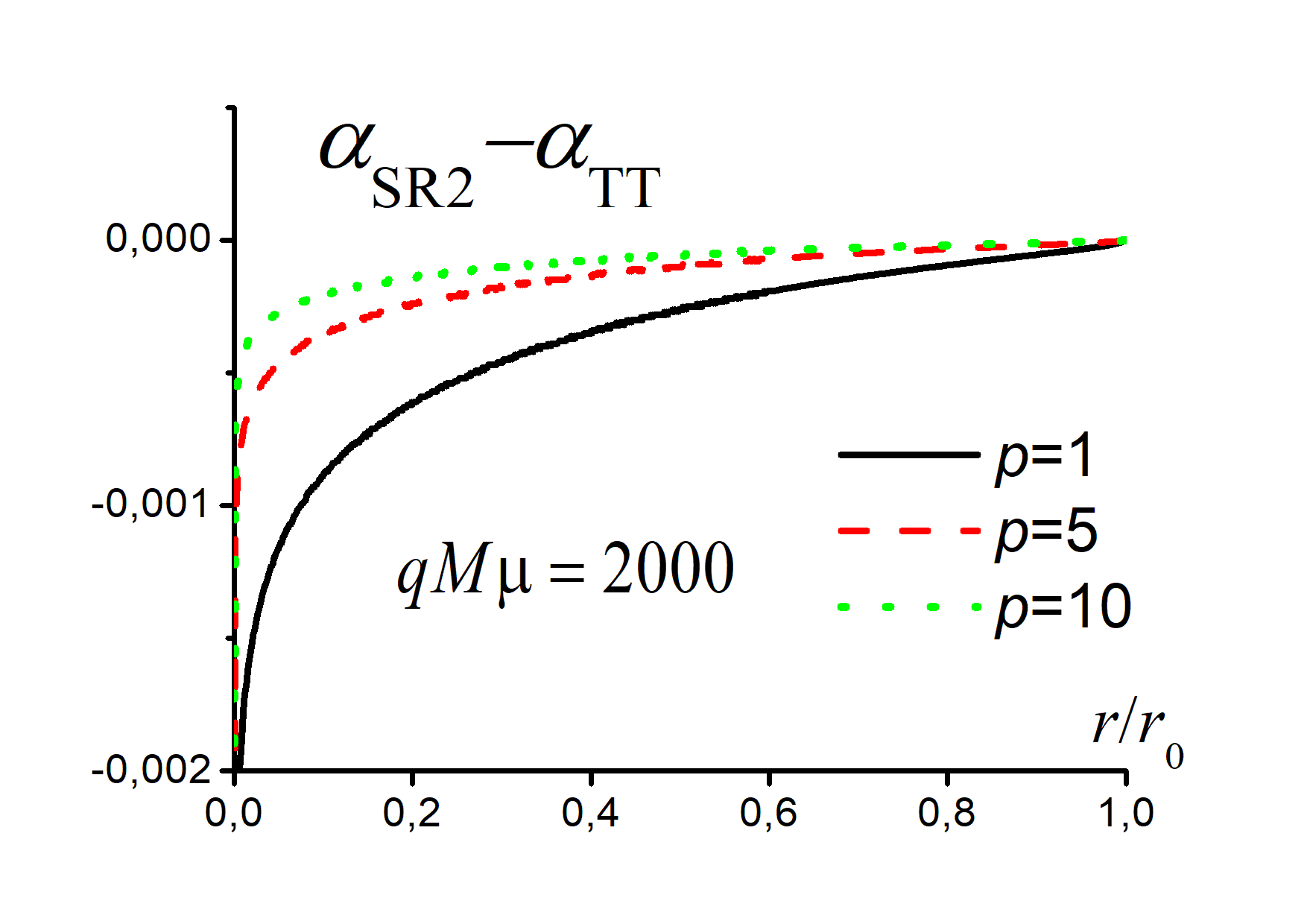}
 	\caption{Difference of $\alpha(r)$ in the scalarization region for  SR2 and TT models with different parameters $p$ for fixed $qM\mu=2000$. The difference practically tends to zero as $p$ increases.}
 	\label{fig1ab:differencesSt-TT}
 \end{figure}
 \begin{figure} [h!]  	\centering
 	\includegraphics[width=0.35 \textwidth]{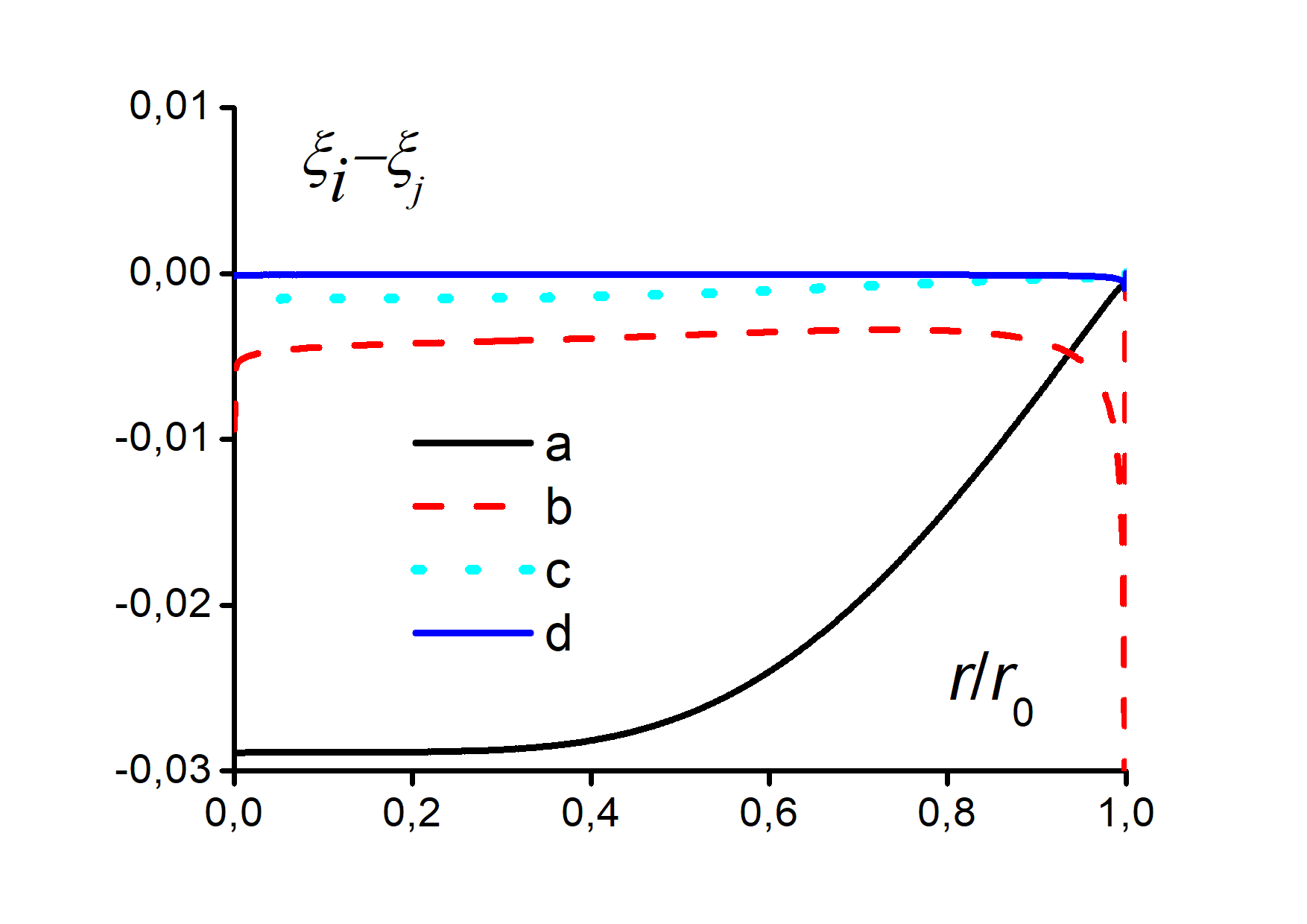}
 	\caption{Differences $\xi_i-\xi_j$  among  models in the scalarization region. Here $\xi_j$ denotes the SR2 model with $M\mu=10^{15}$ and $q=10^5$. The first term, $\xi_i$ corresponds to (a): HT model with $M\mu=1000$, $q=100$ (black solid line); (b): TT model with $M\mu=20$, $q=100, p=20$ (red dashed); (c) TT with  $M\mu=1000$, $q=100, p=50$ (cyan dots); (d): blue solid line (practically zero) corresponds to  the SR2 model with $M\mu=1000$, $q=100$ showing a good approximation to large $M\mu$ limit. }
 	\label{differences_xi}
 \end{figure}
 
 \begin{table}[h]
 	\begin{tabular}{|c|c|c|c|c|}
 		\hline
 	$M\mu$	& $1$ & $100$ & $10^{5}$ & $10^{8}$\\ \hline
 		
 	SR2    & $-6.0\cdot10^{-3}$  &  $-5.9\cdot10^{-5}$   & $-5.9\cdot10^{-8}$ & $-8.7\cdot 10^{-11}$\\ \hline
 	HT  & $-8.2\cdot10^{-3}$  &  $-8.1\cdot10^{-5}$  & $-8.1\cdot10^{-8}$  &  $-1.1\cdot 10^{-10}$   \\ \hline
 	TT, p=5  & $-6.5\cdot 10^{-3}$  &  $-6.4\cdot10^{-5}$  &$-6.4\cdot10^{-8}$&$-9.7\cdot 10^{-11}$ \\ \hline
 	TT, p=50  & $-6.1\cdot 10^{-3}$  &  $-6.0\cdot10^{-5}$  & $-6.0\cdot10^{-8}$ &$-6.6\cdot 10^{-11}$ \\ \hline
 	\end{tabular}
 	\caption{The table shows the value of $\zeta-1$ for different models and $M\mu$ parameters ($q=100$), it illustrates how  $\zeta \to 1$ as $M\mu$ increases, according to (\ref{LimZeta}).  }
 \end{table}
 
 \section{Spherical perturbations}\label{SS_perturbations}
  Here we discuss the stability  of the SSS background against  spherical (radial) perturbations making use of  the basic calculation  performed in \cite{ZhdStSht2024}.  In the  diagonal metric (\ref{metric})  we change  change   $\alpha\to \alpha(x)+\alpha_1(t,x)$, $\beta\to\beta(x)+\beta_1(t,x)$,  $\xi\to\xi(x)+\xi_1(t,x)$, where  $\alpha_1$, $\beta_1$, $\xi_1$ represent small perturbations;  $ \alpha(x)$, $  \beta(x)$, $ \xi(x)$ are the SSS  solutions of the system (\ref{Ein_1-0}), (\ref{Ein_2-0}), \ref{equation-xi-of-x} for the  asymptotically flat  background. 
 Linearization of the Einstein equations with respect to  $\alpha_1$, $\beta_1$, $\xi_1$ leads to one master equation for $ \xi_1$. Here we use a different representation of this equation than that given in \cite{ZhdStSht2024}:
 \begin{equation}
 \frac{A(x)}{\mu^2}\frac{\partial^2\xi_1}{\partial t ^2} =  \frac{\partial }{\partial x}\left[ B(x) \frac{\partial \xi_1}{\partial x}\right]- V_{\rm eff} \xi_1 \,,
 \label{master-radial0_t}
 \end{equation} 
 where  
 \begin{equation}
 A(x)=x^2 e^{\frac{\beta -\alpha }{2}}\,,\quad B(x)=x^2 e^{\frac{\alpha -\beta }{2}}\,,
 \end{equation}
 \begin{align} 	
& V_{\rm eff}(x)=   6x^2e^{\frac{\alpha-\beta}{2}}\frac{d\xi}{dx} \frac{d }{dx}\left(x\frac{d\xi}{dx}\right)\nonumber \\ {}& +x^2e^{\chi}\left[ w'(\xi)x\frac{d\xi}{dx}+\frac{1}{6}w''(\xi)\right] \,,\quad \xi\equiv \xi(x).
 \label{Veff}
 \end{align} 
 Using equations (\ref{dxids}) and (\ref{dZds}) in the form 
 \begin{equation*}
x\frac{d\xi}{dx}=-\frac{Z}{Y}\,, \quad \frac{dZ}{dx}=-\frac16 e^\chi U^2 w'(\xi)\,,
 \end{equation*}
 we have $V_{\rm eff}(x)= X_0^2 e^\chi V^*_{\rm eff}$ , 
 \begin{equation} 	
V^*_{\rm eff}\approx  6  \left(\frac{Z}{Y}\right)^2  U^2 w'(\xi)     +U^2  \left[- 2\frac{Z}{Y}w'(\xi) +\frac{1}{6}w''(\xi)\right] .
 \label{Veff**}
 \end{equation} 
\textbf{The positivity of the effective potential $V_{\rm eff}$ plays a key role. Obviously, the sign of $V^*_{\rm eff}$ (and therefore  $V_{\rm eff}$) is defined by those components  $\xi,Y,Z$ that turn out to be practically unchangeable for sufficiently large $M\mu$. This allows us to extend some of the results of \cite{ZhdStSht2024} to larger values of $M\mu$.} 

In view of (\ref{asympt_r}) $A(x)\sim x^3$,  
 $B(x)\sim x$ and $V_{\rm eff(x)}\to 0$ for $x\to 0$. For $x>X_0$ $V^*_{\rm eff}\approx (x/X_0)^2$
 Equation (\ref{master-radial0_t}) yields
 \begin{align} \label{conservation_1}
 \frac{\partial e}{\partial t}     =  \frac{\partial F  }{\partial x} \,,\quad  F(t,x)= B(x) \frac{\partial \xi_1}{\partial t} \frac{\partial \xi_1}{\partial x} \,,
 \end{align}
  \begin{equation}
 e(t,x)=\frac12 \left[\frac{A}{\mu^2}\left(\frac{\partial \xi_1}{\partial t}\right)^2+ B\left(\frac{\partial \xi_1}{\partial x}\right)^2+V_{\rm eff}\xi_1^2\right]\,.
 \end{equation}

\begin{figure} [h!]  	\centering
  	\includegraphics[width=0.4 \textwidth]{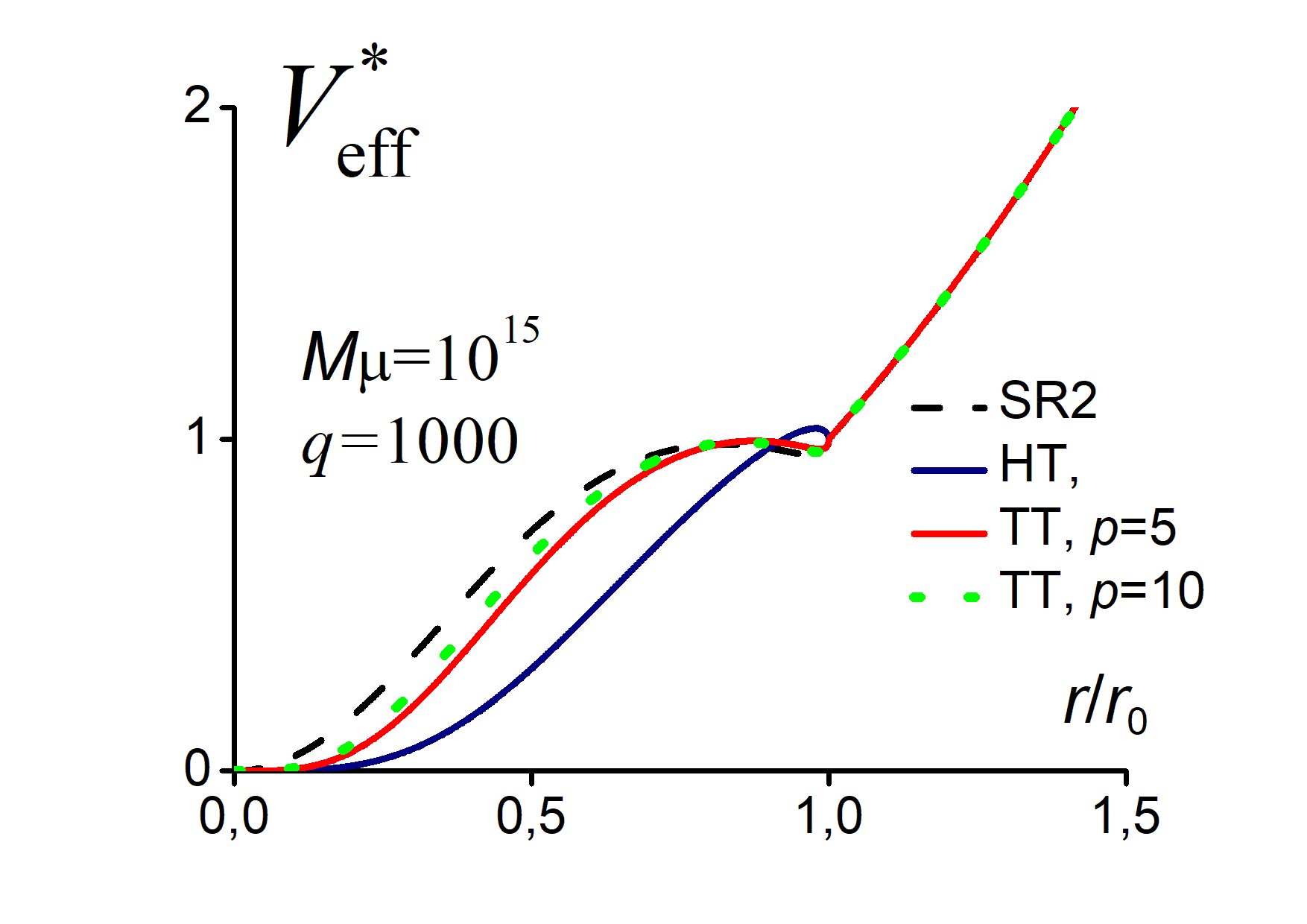}
  	\caption{Effective potential $V^*_{\rm eff}(s)\ge 0 $ for spherical perturbations in case of models SR2, HT, TT ($p=5,10$). In all these models  $M\mu=10^{15}$, $q=1000$. Numerical investigation shows that the graph is practically unchanged for all   $M\mu=100\div 10^{15}$ and $q\sim 10\div 1000$.}
  	\label{FigVeff}
  \end{figure}

 We impose a regular boundary condition by putting
 \begin{equation}\label{b-conditions}
 \lim\limits_{x\to 0+0} F(t,x)=0\,,\quad \lim\limits_{x\to \infty} F(t,x)=0\,.
 \end{equation}
Under this condition, we have   
 \begin{equation}\label{conservation2}
 E(t)=  \int\limits_0^\infty dx\,\, e(t,x) =const\,,
 \end{equation}
The functions $V^*_{\rm eff}$   shown in Fig. \ref{FigVeff} are positive for SR2, HT and TT cases.  Then conservation of $E(t)$ (\ref{conservation2})
   rules out  unstable modes and confirms the  linear  stability of SSS objects in question to radial perturbations for all these models. However, as in work \cite{ZhdStSht2024}, this does not exhaust the issue of stability. To obtain a complete answer,  it is also necessary to investigate polar and axial perturbations.

\bibliography{universality_rev.bib}

\end{document}